\documentclass[10pt,a4paper]{article}
\usepackage[english]{babel}
\usepackage[latin1]{inputenc}
\usepackage{amsfonts,amsbsy,bm,euscript,mathrsfs}
\usepackage{amssymb,stmaryrd,faktor}
\usepackage[tbtags]{amsmath}
\usepackage{bbm}
\usepackage{graphicx}
\usepackage[title,titletoc]{appendix}
\usepackage[bookmarks=true,colorlinks=true,linkcolor=blue,citecolor=blue,urlcolor=blue,bookmarksnumbered]{hyperref}

\usepackage{dsfont}
\usepackage{collref}
\usepackage{lmodern}
\usepackage{mathrsfs}
\usepackage{mathtools}
\usepackage{bbm}
\usepackage{braket}
\usepackage{slashed}

\usepackage{graphicx}
\usepackage{booktabs}
\usepackage{subfig}
\usepackage{tikz}
\usetikzlibrary{plotmarks,calc,decorations,decorations.pathmorphing}

\numberwithin{equation}{section}

\makeatletter
\renewcommand\section{\@startsection {section}{1}{\z@}
{-3.5ex \@plus -1ex \@minus -.2ex}
{2.3ex \@plus.2ex}
{\normalfont\Large\bfseries}}
\renewcommand\subsection{\@startsection{subsection}{2}{\z@}
{-3.25ex\@plus -1ex \@minus -.2ex}
{1.5ex \@plus.2ex}
{\normalfont\large\bfseries}}
\makeatother

\newcommand{\arXivlink}[1]{\href{http://arXiv.org/abs/#1}{arXiv:#1}}

\newcommand{\alg}[1]{\mathfrak{#1}}

\usepackage{blkarray}

\begin{document}
\thispagestyle{empty}
\begin{flushright}\footnotesize\ttfamily
DMUS-MP-23/17
\end{flushright}
\vspace{2em}

\begin{center}

{\Large\bf \vspace{0.2cm}
{\color{black} \large A study of form factors in relativistic mixed-flux $AdS_3$}} 
\vspace{1.5cm}

\textrm{Alessandro Torrielli$^{a}$ \footnote[1]{\textit{E-mail address:} \texttt{a.torrielli@surrey.ac.uk}}}
\vspace{0.8cm}
\\
$^a$ \textit{School  of Mathematics and Physics, University of Surrey, Guildford, GU2 7XH, UK}
\\




\end{center}

\vspace{2em}

\begin{abstract}\noindent 
We study the two-particle form-factors for the relativistic limit of the integrable $S$-matrix of the mixed-flux $AdS_3 \times S^3
 \times T^
4$ string theory. The $S$-matrix theory was formally constructed in two distinct ways by two different teams. We focus on the massive theory built up by Frolov, Polvara and Sfondrini, and derive expressions for the minimal
solutions to the axioms, in both integral and manifestly meromorphic form, and then proceed to apply the off-shell Bethe ansatz method of Babujian {\it et al}. We obtain the integral formulas for the two-particle complete form-factors and check the axioms at this particle number. 
\end{abstract}

\newpage

\overfullrule=0pt
\parskip=2pt
\parindent=12pt
\headheight=0.0in \headsep=0.0in \topmargin=0.0in \oddsidemargin=0in

\vspace{-3cm}
\thispagestyle{empty}
\vspace{-1cm}

\tableofcontents

\setcounter{footnote}{0}
\section{Introduction}

The integrable system underlying string theory in $AdS_3 \times S^3 \times M^4$ \cite{Bogdan}, the manifold $M^4$ either being $T^4$ or $S^3 \times S^1$ \cite{Sundin:2012gc} - see the reviews \cite{rev3,Borsato:2016hud} - is similar but arguably richer than the $AdS_5$ counterpart \cite{Foundations,Beisertreview}. An incomplete list of references is given by \cite{OhlssonSax:2011ms,seealso3,Borsato:2012ss,Borsato:2013qpa,Borsato:2013hoa,PerLinus,Beccaria:2012kb,Sundin:2013ypa,Bianchi:2013nra,Bianchi:2013nra1,Bianchi:2013nra2,CompleteT4,Borsato:2015mma}. In $AdS_3$ massless modes are present \cite{Sax:2012jv,Sax:2014mea,Borsato:2016xns,Baggio:2017kza} and the machinery of massless integrability is called into play \cite{Zamol2,Fendley:1993jh,DiegoBogdanAle}. The possibility of having a mixture of RR and NS-NS flux is also there - in \cite{Bogdan} the case of equal radius for the $S^3 \times S^3 \times S^1$ is discussed and in the generic radius case is found in \cite{OhlssonSax:2011ms}, with the mixed flux case featured in \cite{Ben,Cagnazzo:2012se,Closed}. The paper \cite{Closed} in particular describes the full moduli space, and explains the $h$ parameter which connects with the complete moduli dependence. Recently, the Quantum Spectral Curve (QSC) has been written down for purely RR background $AdS_3 \times S^3 \times T^4$ in \cite{QSC}, see also \cite{Andrea}. A new series of articles \cite{AleSSergey}, see also \cite{Seibold:2022mgg}, has revisited the integrability programme in $AdS_3 \times S^3 \times T^4$, has proposed new dressing phases and has formulated the Thermodynamic Bethe Ansatz (TBA). Further recent work can be found in \cite{further}.


In \cite{gamma1,gamma2} a change of variables was exhibited which shows the complete difference form of the $S$-matrix of massless $AdS_3$ particles. This form coincides with the BMN limit for particles with equal worldsheet chirality \cite{DiegoBogdanAle}. These new (`pseudo-relativistic') massless variables prefigured in \cite{BHL} then led to the massive counterpart being introduced in \cite{AleSSergey}, by virtue of which a part of the massive $S$-matrix (not the whole of it) also acquires a difference form. 

The single-particle mixed-flux dispersion relation is \cite{ArkadyBenStepanchuk,Lloyd:2014bsa} given by
\begin{equation}
\label{eq:disp-rel}
E = \sqrt{\Big(m + \frac{k}{2 \pi} p\Big)^2 + 4 h^2 \sin^2\frac{p}{2}},
\end{equation}
$k \in \mathbbmss{N}$ being the WZW level, $m$ the mass and $h$ the coupling constant. In \cite{gamma2} the massive mixed-flux $S$-matrix was subjected to a particular construction leading to a particular massless relativistic $S$-matrix, for a critical theory whose spectrum is encoded in a massless TBA {\it a la} Zamolodchikov. The paper \cite{Frolov:2023lwd} then obtained the same $S$-matrix from a different limit performed in a way to retain massive excitations, and extended the study to all the different possible masses and bound states. The paper \cite{BOD} provides an in-depth study of the kinematical and analytic structure underlying the theory and its $S$-matrices, and a proposal for the complete mixed-flux dressing factor. More very recent work can be found in \cite{Recent}, as this area is shown to be extremely vibrant.

In this paper we wish to initiate the study of relativistic form factors for this massive theory. The form factor programme is very much at the core of $AdS_5/CFT_4$ in the shape of the hexagon approach \cite{Basso:2013vsa}. This approach has been generalised to $AdS_3$ \cite{Eden:2021xhe}. The traditional relativistic form factor programme was initially pursued in \cite{Thomas}, where it was extended to non-difference form. In \cite{AleForm} the standard ({\it pre-hexagon}) approach was applied to massless $AdS_3$ in the BMN-limit, which is a genuinely relativistic model \cite{DiegoBogdanAle} - see also \cite{gamma2,AleSSergey,Bielli:2023gke} - by following the off-shell Bethe Ansatz technique of \cite{BabuF}. We now intend to do the same for the simplest form factors of this deformed system at hand.

\section{Form factors}

The study of form factors is essential to bring to completion the {\it bootstrap programme} and reach a complete solution of an integrable quantum field theory \cite{Babu,DelfinoMussSimo,fofa}. After having established the exact $S$-matrix and the spectrum of (bound) states, one proceeds with the form factor analysis \cite{Karo1}. Using the information contained in the form factors of the theory, one can then compute the $n$-point correlation functions. This effectively defines the theory, even in those situation where one does not have access to a Lagrangian. We refer the reader to \cite{Babu} for a very thorough review of the form factor programme. 

In a relativistic theory the $n$-particle form factor associated with a operator $\cal{O}$ sitting at the spatial origin is defined as the quantity
\begin{equation}
\label{in}
F^{\cal{O}}_{\alpha_1 ... \alpha_n} (\theta_1,...,\theta_n) = \langle 0| {\cal{O}}(0)|\theta_1,...,\theta_n\rangle_{\alpha_1 ... \alpha_n},
\end{equation}
$\theta_i$ being the rapidity and $\alpha_i$ collectively labelling the quantum numbers of the $i$-th magnon. 

The form factors are constrained by a list of axioms - we refer to \cite{Babu} and \cite{BabuF} for a complete description. We shall only confine ourselves here to those axioms which will be important in our calculation. We have
\begin{itemize}

\item {\it Permutation}
\begin{eqnarray}
&&F^{\cal{O}}_{\alpha_1 ... \alpha_{j-1} \, \beta_{j} \, \beta_{j+1} \, \alpha_{j+2} ... \alpha_n} (\theta_1,...,\theta_{j-1} , \theta_{j} , \theta_{j+1} , \theta_{j+2}, ... \theta_n) = \\
&&\qquad \qquad F^{\cal{O}}_{\alpha_1 ... \alpha_{j-1} \, \alpha_j \, \alpha_{j+1} \, \alpha_{j+2} ... \alpha_n} (\theta_1,...,\theta_{j-1} , \theta_{j+1} , \theta_j, \theta_{j+2}, ... \theta_n) \, S^{\alpha_j \alpha_{j+1}}_{\beta_j \beta_{j+1}} (\theta_j - \theta_{j+1}),\nonumber
\end{eqnarray}
where we adopt the conventions of \cite{BabuF}.
\item {\it Periodicity}
\begin{eqnarray}
&&F^{\cal{O}}_{\alpha_1 \, \alpha_{2} ... \alpha_{n-1} \, \alpha_n} (\theta_1 + 2i \pi,\theta_{2}, ... \theta_{n-1}, \theta_n) = (-)^\sigma\, F^{\cal{O}}_{\alpha_2 \, \alpha_{3} ... \alpha_{n} \, \alpha_1} (\theta_2 , \theta_{3}, ... \theta_{n} , \theta_1),
\end{eqnarray}
where $\sigma$ is an appropriate statistical factor.

\item {\it Lorentz boost}
\begin{eqnarray}
F^{\cal{O}}_{\alpha_1 \, \alpha_{2} ... \alpha_{n-1} \, \alpha_n} (\theta_1 + \Lambda,\theta_{2}+\Lambda, ... \theta_{n-1}+\Lambda, \theta_n+\Lambda) = e^{s \Lambda}F^{\cal{O}}_{\alpha_1 \, \alpha_{2} ... \alpha_{n-1} \, \alpha_n} (\theta_1,\theta_{2}, ... \theta_{n-1}, \theta_n), 
\end{eqnarray}
where $s$ denotes the spin of $\cal{O}$.

\end{itemize}

We will comment later on on the kinematical singularity and bound state singularity axioms.

The method which we will follow in this paper to obtain the two-particle form factor is the off-shell Bethe ansatz \cite{BabuF}. We will describe in detail how this works. Here we provide a short summary of the building blocks which will enter the off-shell Bethe ansatz, and how we will assemble them, to guide with the organisation of the material. This structure follows the decomposition of the $S$-matrix as \cite{Frolov:2023lwd}
\begin{eqnarray}
S(m_1,m_2;\theta) = \Phi(m_1,m_2;\theta) \, \sigma^{-2}(m_1,m_2;\theta) \, S^{BB}(m_1,m_2;\theta) \hat{\otimes} S^{BB}(m_1,m_2;\theta),
\end{eqnarray}
where $\Phi$ and $\sigma^{-2}$ are scalar factors, and $S^{BB}$ is the $S$-matrix whose associated $R$-matrix will be $R(\theta)$ (\ref{rma}) - with $\hat{\otimes}$ being the tensor product of two copies of the fundamental representation to be discussed around (\ref{tenso}), the hat just helping to keep it separated from $\otimes$ which denotes the tensor product of the scattering particles. The integers $m_1$ and $m_2$ associated with the two particles are quantum numbers which enter all the scattering data, the particle mass being given by
\begin{eqnarray}
mass = 2 \, \Big\vert \sin \frac{\pi m}{k} \Big\vert.
\end{eqnarray} 

Accordingly, the two-particle form factor is build out of: 

\begin{itemize}

\item a block associated with $\Phi$, which will be called $F_{pre}^{total}(m_1,m_2;\theta)$ as in ``pre-factor";

\item a block associated with $\sigma^{-2}$, which will be called $F^{dressing}(m_1,m_2;\theta)$ as in ``dressing factor" to adopt the nomenclature of \cite{Frolov:2023lwd}; 

\item a block associated with $S^{BB} \hat{\otimes}S^{BB}$, made out of two copies of the sub-block to be called $G_{ab}(\theta_1,\theta_2)$ - this sub-block will have two indices (each index carrying the fundamental representation) and it will depend specifically on the two scattering rapidities $\theta_1$ and $\theta_2$ to incorporate operators with spin.   

\end{itemize}  

To construct these blocks we shall need first to build the minimal solutions and study their singularities. We then present the core of the off-shell Bethe ansatz and proceed to reconstitute the complete list of blocks. The blocks will eventually be combined as
\begin{eqnarray}
F_{pre}^{total}(m_1,m_2;\theta) F^{dressing}(m_1,m_2;\theta) G_{ac}(\theta_1,\theta_2)G_{bd}(\theta_1,\theta_2)
\end{eqnarray}
to form the complete two-particle form factor. 

\section{Minimal two-particle form factor}

We can calculate the minimal two-particle form factor by using a classic result by Karowski and Weisz \cite{Karo1}. We start from the explicit meromorphic expression for the dressing factor \cite{Frolov:2023lwd}\footnote{An alternative formula for this dressing function, numerically coincident with (\ref{ratio}) in a restricted domain, has been constructed in \cite{gamma2} for the special case of what would be here $m_1=m_2=1$.}
\begin{eqnarray}
\sigma^{-2}(m_1,m_2;\theta) = \frac{R_0^2(\theta - i m_+)\, R_0^2(\theta + i m_+)}{R_0^2(\theta - i m_-)\, R_0^2(\theta + i m_-)}, \qquad m_\pm = \frac{\pi}{k} (m_1 \pm m_2),\label{ratio}
\end{eqnarray}
where
\begin{eqnarray}
R_0(\theta) = \Big(\frac{e}{2 \pi}\Big)^{\tau} \prod_{\ell=1}^\infty \frac{\Gamma(\ell + \tau)}{\Gamma(\ell - \tau)}e^{-2\tau \psi(\ell)}, \qquad \tau = \frac{\theta}{2\pi i},\label{parts}
\end{eqnarray}
$\psi$ being the Digamma function $\psi(z) = \frac{d}{dz}\log \Gamma(z)$. In the ratio (\ref{ratio}) only the Gamma functions, of all the parts of (\ref{parts}), do not cancel out, and one gets
as a net result 
\begin{eqnarray}
\sigma^{-2}(m_1,m_2;\theta) = \prod_{\ell=1}^\infty \frac{\Gamma^2(\ell + \tau - i \mu_+)\Gamma^2(\ell + \tau + i \mu_+)\Gamma^2(\ell - \tau + i \mu_-)\Gamma^2(\ell - \tau - i \mu_-)}{\Gamma^2(\ell - \tau + i \mu_+)\Gamma^2(\ell - \tau - i \mu_+)\Gamma^2(\ell + \tau - i \mu_-)\Gamma^2(\ell + \tau + i \mu_-)},\label{nume}
\end{eqnarray}
where
\begin{eqnarray}
\mu_\pm = \frac{m_\pm}{2\pi i}.
\end{eqnarray}
The infinite product (\ref{nume}) converges to a meromorphic function in the whole complex plane of $\theta$, thanks to the subtle balance of the arguments of the Gamma functions in the numerator and denominator \cite{Kulkarni}.
The dressing factor satisfies
\begin{eqnarray}
\sigma^{-2}(m_1,m_2;\theta)\sigma^{-2}(m_2,m_1;-\theta)=1,
\end{eqnarray} 
since $\mu_\pm \to \pm \mu_\pm$ if we exchange $m_1 \leftrightarrow m_2$.

We now adopt the standard technique of resorting to the Malmst\'en representation of the Gamma function:
\begin{eqnarray}
\Gamma(z) = \exp \int_0^\infty \frac{dx}{x}e^{-x} \Big[z-1-\frac{1-e^{-x(z-1)}}{1-e^{-x}}\Big].\label{Mal}
\end{eqnarray}
Thanks to the fact that the sum of the arguments of all the Gamma functions at the numerator of (\ref{nume}) equals the sum of the arguments at the denominator, we see that most of (\ref{Mal}) cancels out in the ratio, and only the term containing $e^{-x(z-1)}$ contributes. At this point, the product over $\ell$ becomes a simple geometric sum. Collecting all such terms, and rescaling the integration variable as $x \to 2x$, we obtain  
\begin{eqnarray}
\sigma^{-2}(m_1,m_2;\theta) = \exp \int_0^\infty \frac{dx}{x} \frac{8 \, e^{2x}}{\big(e^{2x}-1 \big)^2}\Big[\cosh x M_- - \cosh x M_+\Big] \sinh \frac{x \theta}{i \pi}, \qquad M_\pm = \frac{m_1 \pm m_2}{k}. \label{numero}
\end{eqnarray}
We can clearly see the interesting property of the dressing factor to go into its inverse as we exchange $M_+ \leftrightarrow M_-$.
The expression (\ref{numero}) is exactly of the Karowski-Weisz form
\begin{eqnarray}
\sigma^{-2}(m_1,m_2;\theta) = \exp \int_0^\infty dx \, f(x) \, \sin \frac{x \theta}{i \pi}, \label{Karo0}
\end{eqnarray}
 and therefore we can immediately find the minimal two-particle form factor:
\begin{eqnarray}
F^{-2}(m_1,m_2;\theta) = \exp \int_0^\infty dx \, f(x) \, \frac{\sin^2 \frac{x(i\pi - \theta)}{2\pi}}{\sinh x}, \label{Karo}
\end{eqnarray}
where
\begin{eqnarray}
f(x) = \frac{8 \, e^{2x}}{x\big(e^{2x}-1 \big)^2}\Big[\cosh x M_- - \cosh x M_+\Big].
\end{eqnarray}

By carefully reconstructing backwards a process similar to the one which we followed when going from (\ref{nume}) to (\ref{numero}), it is possible to find the analytic continuation of (\ref{Karo}) to the entire complex plane of $\theta$. We summarise the main steps here, while we refer to \cite{AleForm} - section 3.2 in that paper - for a more detailed procedure applied to the case of a completely analogous calculation. It is sufficient to i) rescale the variable $x \to \frac{x}{2}$; ii) use two of the three factors $(e^x-1)$ at the denominator (one coming from the $\sinh x$) to reconstruct two geometric sums, the third factor eventually reconstructing the denominator of (\ref{Mal}); iii) expand the numerator into exponentials and check that the sum of arguments balance between exponentials with the plus (which will end up in the Gamma functions at the numerator) and exponentials with the minus (which will end up in the Gamma functions at the denominator). In this way we have reconstructed the Malmst\'en representation proceeding backwards, and we find the following result:
\begin{eqnarray}
&&F^{-2}(m_1,m_2;\theta) = \prod_{m,\ell=1}^\infty \frac{\Gamma^2\big(m+\ell-\tau -\frac{M_+}{2}\big)\Gamma^2\big(m+\ell-\tau +\frac{M_+}{2}\big)}{\Gamma^2\big(m+\ell-\tau -\frac{M_-}{2}\big)\Gamma^2\big(m+\ell-\tau +\frac{M_-}{2}\big)} \times \nonumber\\
&&\qquad \qquad \qquad \times \frac{\Gamma^2\big(m+\ell+\tau -\frac{M_+}{2}-1\big)\Gamma^2\big(m+\ell+\tau +\frac{M_+}{2}-1\big)}{\Gamma^2\big(m+\ell+\tau -\frac{M_-}{2}-1\big)\Gamma^2\big(m+\ell+\tau +\frac{M_-}{2}-1\big)} \times\nonumber\\
&&\qquad \qquad \qquad \times \frac{\Gamma^4\big(m+\ell-\frac{M_-}{2}-\frac{1}{2}\big)\Gamma^4\big(m+\ell+\frac{M_-}{2}-\frac{1}{2}\big)}{\Gamma^4\big(m+\ell -\frac{M_+}{2}-\frac{1}{2}\big)\Gamma^4\big(m+\ell +\frac{M_+}{2}-\frac{1}{2}\big)}.\label{third}
\end{eqnarray}
This product converges for the already mentioned delicate balance of arguments of the Gamma functions \cite{Kulkarni}. 
We can still see the interesting property of $F(m_1,m_2;\theta)$ to go into its inverse as we exchange $M_+ \leftrightarrow M_-$. The formula (\ref{third}) is valid for sufficiently generic values of $M_\pm = \frac{m_1\pm m_2}{k}$. For specific values of $m_1,m_2$ and $k$ there may be accidental singularities in the third line of (\ref{third}), which does not contain the variable $\tau = \frac{\theta}{2 \pi i}$: these cases require to be dealt with separately and may potentially be subject to drastic simplifications.

By the Karowski-Weisz result, combined with the fact that both $\sigma$ and $F$ are symmetric under the exchange $m_1 \leftrightarrow m_2$, we conclude that 
\begin{eqnarray}
F^{-2}(m_1,m_2;\theta) = F^{-2}(m_2,m_1;-\theta) \sigma^{-2}(m_1,m_2;\theta), \qquad F^{-2}(m_1,m_2;i\pi - \theta) = F^{-2}(m_1,m_2;i\pi + \theta).\nonumber
\end{eqnarray}

\subsection{Singularities of $\sigma^{-2}(m_1,m_2;\theta)$ and $F^{-2}(m_1,m_2;\theta)$}

For future purposes, let us list the poles and zeros of $\sigma^{-2}(m_1,m_2;\theta)$, as can be evinced from (\ref{nume}). We list the labels first in a rather pleonastic way - one can in fact streamline the counting. 

\begin{itemize}

\item Poles: if $\ell = 1,2,3,..$ and $n=0,1,2,3,...$, we have
\begin{eqnarray}
&&\tau = i \mu_+ - (n+\ell) = i\mu_+ - q, \qquad q = 1,2,3,..., \qquad \mbox{pole of order} \, \,  2q,\nonumber\\
&&\tau = -i \mu_+ - (n+\ell) = -i\mu_+ - q, \qquad q = 1,2,3,..., \qquad \mbox{pole of order} \, \,  2q,\nonumber\\&&\tau = i \mu_- + (n+\ell) = i\mu_- + q, \qquad q = 1,2,3,..., \qquad \mbox{pole of order} \, \,  2q,\nonumber\\&&\tau = -i \mu_- + (n+\ell) = -i\mu_- + q, \qquad q = 1,2,3,..., \qquad \mbox{pole of order} \, \,  2q.
\end{eqnarray}

\item Zeros: if $\ell = 1,2,3,..$ and $n=0,1,2,3,...$, we have
\begin{eqnarray}
&&\tau = i \mu_+ + (n+\ell) = i\mu_+ + q, \qquad q = 1,2,3,..., \qquad \mbox{zero of order} \, \,  2q,\nonumber\\
&&\tau = -i \mu_+ + (n+\ell) = -i\mu_+ + q, \qquad q = 1,2,3,..., \qquad \mbox{zero of order} \, \,  2q,\nonumber\\&&\tau = i \mu_- - (n+\ell) = i\mu_- - q, \qquad q = 1,2,3,..., \qquad \mbox{zero of order} \, \,  2q,\nonumber\\&&\tau = -i \mu_- - (n+\ell) = -i\mu_- - q, \qquad q = 1,2,3,..., \qquad \mbox{zero of order} \, \,  2q.
\end{eqnarray}
\end{itemize}

We recall that
\begin{eqnarray}
i \mu_\pm = \frac{m_1 \pm m_2}{2k}.
\end{eqnarray}
All these values are therefore located on the imaginary $\theta$ axis. Generically there is no cancellation between poles and zeros, although there may be cancellations for special values of $m_1$ and $m_2$.

Likewise, the singularities of $F^{-2}(m_1,m_2;\theta)$ can be evinced from (\ref{third}).
\begin{itemize}

\item Poles: for $m,\ell = 1,2,3,...$ and $n=0,1,2,3,...$, we have
\begin{eqnarray}
&&\tau = m+\ell +n -\frac{M_+}{2}= -\frac{M_+}{2} +q, \qquad q=2,3,..., \qquad \mbox{pole of order two, six,....}\nonumber\\
&&\tau = m+\ell +n +\frac{M_+}{2}= \frac{M_+}{2} +q, \qquad q=2,3,..., \qquad \mbox{pole of order two, six,....}\nonumber\\
&&\tau = -m-\ell -n +\frac{M_+}{2}= +\frac{M_+}{2} -q, \qquad q=1,2,..., \qquad \mbox{pole of order two, six,....}\nonumber\\
&&\tau = -m-\ell -n -\frac{M_+}{2}= -\frac{M_+}{2} -q, \qquad q=1,2,..., \qquad \mbox{pole of order two, six,...}.\nonumber\\
\end{eqnarray}

\item Zeros: for $m,\ell = 1,2,3,...$ and $n=0,1,2,3,...$, we have
\begin{eqnarray}
&&\tau = m+\ell +n -\frac{M_-}{2}= -\frac{M_-}{2} +q, \qquad q=2,3,..., \qquad \mbox{zero of order two, six,....}\nonumber\\
&&\tau = m+\ell +n +\frac{M_-}{2}= \frac{M_-}{2} +q, \qquad q=2,3,..., \qquad \mbox{zero of order two, six,....}\nonumber\\
&&\tau = -m-\ell -n +\frac{M_-}{2}= +\frac{M_-}{2} -q, \qquad q=1,2,..., \qquad \mbox{zero of order two, six,....}\nonumber\\
&&\tau = -m-\ell -n -\frac{M_-}{2}= -\frac{M_-}{2} -q, \qquad q=1,2,..., \qquad \mbox{zero of order two, six,...}.\nonumber\\
\end{eqnarray}

\end{itemize}

Let us notice for futures scopes that $F^{-1}$ and $\sigma^{-1}$ are also well-defined and meromorphic, with poles and zeros starting from a minimum order one instead of a minimum order two.

\section{Tools from the algebraic Bethe ansatz}

The minimal form factor is just a building block on the way towards the complete form factor formula. The method of the off-shell Bethe ansatz \cite{BabuF} requires the use of the $\hat{C}$ operator borrowed from the algebraic Bethe ansatz. We do this for two physical spaces (and one auxiliary space). It is important to remark that the integral formulas constructed in \cite{BabuF} for the Sine-Gordon model assume an odd number of particles. We will find here that we can solve the axioms using integral expressions based on the off-shell Bethe ansatz for an even particle number (specifically equal to two) as well. We decided to build integral expressions in this case both i) to prove that they are actually solutions, and ii) to setup the type of building blocks which will be useful in the future in an all-particle generalisation.    

The $R$-matrix associated with the massive-massive scattering is given by \cite{Frolov:2023lwd}\footnote{This $R$-matrix generalises the formula which was used in \cite{gamma2} for $m_1=m_2=1$, where it also coincides with \cite{qseealso}, resp. \cite{Ben}, for $q$-deformed, resp. Pohlmeyer-reduced relativistic theories. We refer to section 3.1 of \cite{gamma2} for a more detailed discussion of these relations.}
\begin{eqnarray}
&&R(\theta) = A(\theta) {\bf E}_{11}\otimes {\bf E}_{11} + B(\theta) {\bf E}_{11}\otimes {\bf E}_{22} + C(\theta) {\bf E}_{21}\otimes {\bf E}_{12} -E(\theta) {\bf E}_{12}\otimes {\bf E}_{21}\nonumber\\
&&\qquad \qquad  \qquad \qquad \qquad \qquad \qquad \qquad \qquad \qquad +D(\theta) {\bf E}_{22}\otimes {\bf E}_{11} -G(\theta) {\bf E}_{22}\otimes {\bf E}_{22},\label{rma}
\end{eqnarray} 
the matrices ${\bf E}_{ij}$ having all zeros except a $1$ in row $i$, column $j$. For  our purposes the state $|1\rangle = |\phi\rangle$ is a boson and the state $|2\rangle = |\psi\rangle$ is a fermion. We will confine ourselves to massive excitations in this paper. The entries are spelt out in (C.4) of \cite{Frolov:2023lwd}, and they read
\begin{eqnarray}
&&A(\theta) = 1, \qquad B(\theta) = \frac{{\cal{S}}_1 e^{\frac{i m_2 \pi}{k}+\theta} - {\cal{S}}_2 e^{\frac{i m_1 \pi}{k}}}{{\cal{S}}_1 e^{\frac{i (m_1+m_2) \pi}{k}+\theta} - {\cal{S}}_2},\nonumber\\
&&C(\theta) = \frac{i e^{\frac{i m_1 \pi}{k}+\frac{\theta}{2}}\sqrt{\sigma(m_1) \sigma(m_2)}}{{\cal{S}}_1 e^{\frac{i (m_1+m_2) \pi}{k}+\theta} - {\cal{S}}_2}, \qquad D(\theta) = \frac{{\cal{S}}_1 e^{\frac{i m_1 \pi}{k}+\theta} - {\cal{S}}_2 e^{\frac{i m_2 \pi}{k}}}{{\cal{S}}_1 e^{\frac{i (m_1+m_2) \pi}{k}+\theta} - {\cal{S}}_2},\nonumber\\
&&E(\theta)=\frac{i e^{\frac{i m_2 \pi}{k}+\frac{\theta}{2}}\sqrt{\sigma(m_1) \sigma(m_2)}}{{\cal{S}}_1 e^{\frac{i (m_1+m_2) \pi}{k}+\theta} - {\cal{S}}_2}, \qquad G(\theta) = \frac{-{\cal{S}}_1 e^{\theta} + {\cal{S}}_2 e^{\frac{i (m_1+m_2) \pi}{k}}}{{\cal{S}}_1 e^{\frac{i (m_1+m_2) \pi}{k}+\theta} - {\cal{S}}_2},
\end{eqnarray}  
where
\begin{eqnarray}
{\cal{S}}_i = \mbox{sgn}\big[\sin \frac{\pi m_i}{k}\big], \qquad m_i \neq 0 \, \, \mbox{mod} \, \, k,
\end{eqnarray}
and
\begin{eqnarray}
\sigma(m) = 2 \, \Big\vert \sin \frac{\pi m}{k}\Big\vert.
\end{eqnarray}
Our conventions are such that, if we do not write it explicitly, we always intend
\begin{eqnarray}
A(\theta) = A(m_1,m_2;\theta), \qquad \mbox{etc.},\qquad \theta = \theta_1-\theta_2,
\end{eqnarray}
while we shall explicitly indicate any other assignment - for instance $A(m_2,m_1;\theta)$.
We shall also need the $S$-matrix, which is obtained as
\begin{eqnarray}
S = \Pi \circ R,
\end{eqnarray}
where $\Pi$ is the graded permutation acting on two states:
\begin{eqnarray}
\Pi \, |v\rangle \otimes |w\rangle = (-)^{deg(v)deg(w)} |w\rangle \otimes |v\rangle,
\end{eqnarray} 
and $deg(\phi)=0$, $deg(\psi)=1.$ 

Even if the Yang-Baxter equation is solved in the general case, will always assume in this paper that ${\cal{S}}_i = 1$, for all $i=0,1,2$ (with the auxiliary space $0$ to appear later on). This is achieved (for the case of non-zero mass which we are concerned with here) by assuming that $m \in \{1,2,...,k-1\}$, which is sufficient to cover the range of the physical massive spectrum \cite{Frolov:2023lwd}.

We construct the monodromy matrix as
\begin{eqnarray}
{\cal{M}} = R_{10}(\theta_1 - \theta_0) R_{20}(\theta_2 - \theta_0),\label{evenif}
\end{eqnarray}
where the indices denote the space the $R$-matrix is acting on. The $\hat{C}$ operator is obtained by extracting the terms with an $E_{21}$ in the auxiliary space. We find
\begin{eqnarray}
\hat{C} = -E_{10}A_{20} \, {\bf E}_{12} \otimes {\bf E}_{11} -E_{10}D_{20} \, {\bf E}_{12} \otimes {\bf E}_{22} -B_{10}E_{20} \, {\bf E}_{11} \otimes {\bf E}_{12} +F_{10}E_{20} \, {\bf E}_{22} \otimes {\bf E}_{12},
\end{eqnarray}
having indicated
\begin{eqnarray}
A_{i0} \equiv A(m_i,m_0;\theta_i-\theta_0), \qquad i=1,2, \qquad \mbox{etc.} 
\end{eqnarray}
We shall need the action on the pseudovacuum covector:
\begin{eqnarray}
\langle \phi \phi |\hat{C} = -E_{10}A_{20} \langle \psi \phi | -B_{10}E_{20} \langle \phi \psi |.\label{numi}
\end{eqnarray}
We shall also need
\begin{eqnarray}
\langle \phi \phi |\hat{C} \hat{C} = - (B_{10}E_{20}E_{10}D_{20}+E_{10}A_{20}F_{10}E_{20}) \langle \psi \psi |.\label{ff}
\end{eqnarray}

\subsection{The off-shell Bethe ansatz}

The off-shell Bethe ansatz is designed to provide a solution to the permutation axiom of form factors. We follow closely the conventions of \cite{BabuF} throughout the paper. 


The $S$-matrix of the theory is given as a product of various pieces - each piece will provide a certain contribution to the actual form factor. All these contributions will have to be multiplied together at the end. 
 
\subsubsection{Prefactor}

There is a contribution from a specific prefactor \cite{Frolov:2023lwd} 
\begin{eqnarray}
\Phi(m_1,m_2;\theta) = \frac{\prod_{n=1}^N \big( \big[|m_1-m_2| + 2n]_\theta\big)^2}{\big[|m_1-m_2| \big]_\theta \big[ m_1 + m_2\big]_\theta},
\end{eqnarray}
with $N = m_1$ if $m_1 \leq m_2$ and $N=m_2$ if $m_2<m_1$, and
where 
\begin{eqnarray}
[m]_\theta \equiv \frac{\sinh \Big(\frac{\theta}{2} + \frac{i \pi m}{2k}\Big)}{\sinh \Big(\frac{\theta}{2} - \frac{i \pi m}{2k}\Big)}.
\end{eqnarray}
The contribution to the complete form factor from each of the individual factors $[m]_\theta$ can be found using the formulas in appendix C of \cite{BabuF}. Namely, every factor
\begin{eqnarray}
-[m]_\theta = \frac{\sin \frac{\pi}{2}(a+\frac{\theta}{i\pi})}{\sin \frac{\pi}{2}(a-\frac{\theta}{i\pi})} = \exp 2 \int_0^\infty \frac{dt}{t} \, \frac{\sinh t (1-a)}{\sinh t} \, \sinh \frac{t \theta}{i\pi}, 
\end{eqnarray} 
where we have set 
\begin{eqnarray}
a = \frac{m}{k},
\end{eqnarray}
contributes a corresponding term
\begin{eqnarray}
&&f_{pre}(\theta,m) = \exp 2 \int_0^\infty \frac{dt}{t} \, \frac{\sinh t (1-a)}{\sinh t} \, \frac{1- \cosh t(1-\frac{\theta}{i\pi})}{\sinh t} =\nonumber\\
&&\qquad \qquad \prod_{n=0}^\infty \frac{\Gamma(n+1-\frac{a}{2} + \tau)\Gamma(n+2-\frac{a}{2} - \tau)\Gamma(n+\frac{1}{2}+\frac{a}{2})^2}{\Gamma(n+\frac{a}{2} + \tau)\Gamma(n+1+\frac{a}{2} - \tau)\Gamma(n+\frac{3}{2}-\frac{a}{2})^2},\qquad a = \frac{m}{k},\label{fpre}
\end{eqnarray}
to the complete form factor.
The total number of factors $[m]_\theta$ is always even so the minus sign does not matter, therefore the total contribution from this part is
\begin{eqnarray}
F^{total}_{pre}(m_1,m_2;\theta) = \frac{\prod_{n=1}^N f_{pre}^2(\theta,|m_1-m_2| + 2n)}{f_{pre}(\theta,|m_1-m_2| ) f_{pre}( \theta,m_1 + m_2)},
\end{eqnarray}
with $N = m_1$ if $m_1 \leq m_2$ and $N=m_2$ if $m_2<m_1$. By the usual Karowski-Weisz theorem, combined with the fact that both $\Phi$ and $F^{total}_{pre}$ are symmetric under the exchange $m_1 \leftrightarrow m_2$, we conclude that
\begin{eqnarray}
&&F^{total}_{pre}(m_1,m_2;\theta) = \Phi(m_1,m_2;\theta) F^{total}_{pre}(m_2,m_1;-\theta) \nonumber\\
&&F^{total}_{pre}(m_1,m_2;i\pi - \theta)=F^{total}_{pre}(m_1,m_2;i \pi+\theta). 
\end{eqnarray}

\subsubsection{Dressing factor}

There is then a contribution from the dressing factor $\sigma^{-2}(m_1,m_2;\theta)$. This will contribute a term $F^{dressing}$ such that
\begin{eqnarray}
F^{dressing}(m_1,m_2;\theta_1,\theta_2) = \sigma^{-2}(m_1,m_2;\theta_1-\theta_2) F^{dressing}(m_2,m_1;\theta_2,\theta_1).
\end{eqnarray}
It is clear that we can simply set
\begin{eqnarray}
F^{dressing}(m_1,m_2;\theta_1,\theta_2) \equiv F^{dressing}(m_1,m_2;\theta_1-\theta_2)= F^{-2}(m_1,m_2;\theta_1-\theta_2),
\end{eqnarray}
where $F^{-2}(m_1,m_2;\theta_1,\theta_2)$ is our minimal form-factor solution (\ref{third}). We then also have therefore
\begin{eqnarray}
F^{dressing}(m_1,m_2;i \pi - \theta) = F^{dressing}(m_1,m_2;i \pi + \theta).
\end{eqnarray}


\subsubsection{Matrix part}

The matrix part of the complete massive $R$-matrix is composed of two copies of the $\mathfrak{psu}(1|1)^2$ $R$-matrix (\ref{rma}). It is not difficult to convince oneself that this implies a factorised structure: we can solve the permutation axiom for each individual copy of $\mathfrak{psu}(1|1)^2$, and multiply the components to obtain the total contribution. More precisely, if we denote a state in the fundamental four-dimensional representation of $\mathfrak{psu}(1|1)^4$ by
\begin{eqnarray}
|a \hat{\otimes} b\rangle,\label{tenso}
\end{eqnarray}
with $a$ and $b$ in the fundamental two-dimensional representation of $\mathfrak{psu}(1|1)^2$,
the complete $R$-matrix $R_{tot}$ will act on two such states as
\begin{eqnarray}
R_{tot} (|a \hat{\otimes} b\rangle)_{\theta_1} \otimes (|c \hat{\otimes} d\rangle)_{\theta_2}
= R_{ac}^{AC} (\theta_1,\theta_2) \, R_{bd}^{BD} (\theta_1,\theta_2) \, (|A \hat{\otimes} B\rangle)_{\theta_1} \otimes (|C \hat{\otimes} D\rangle)_{\theta_2},
\end{eqnarray}
where $R$ is the $4 \times 4$ $R$-matrix (\ref{rma}).
If we split the total contribution of the matrix part as
\begin{eqnarray}
F^{matrix}_{(a \hat{\otimes} b)(c \hat{\otimes} d)}(\theta_1,\theta_2) = G_{ac}(\theta_1,\theta_2) \, G_{bd}(\theta_1,\theta_2),
\end{eqnarray}
then the individual functions $G_{ac}(\theta_1,\theta_2)$ and $G_{bd}(\theta_1,\theta_2)$ satisfy the permutation axiom with respect to the individual $R$-matrix $R$. These functions will now be built as
\begin{eqnarray}
G_{ab}(\theta_1,\theta_2) = {\cal{I}}_0 . F_{ab}(m_0,m_1,m_2;\theta_0,\theta_1,\theta_2),
\end{eqnarray}
where ${\cal{I}}_0$ is an integral operator with respect to an auxiliary integration variable $\theta_0$ which we will define later on, while the $F_{ab}(m_0,m_1,m_2;\theta_0,\theta_1,\theta_2)$ block will come from the off-shell Bethe ansatz.

Let us therefore focus on the individual combinations of states. 

{\it $\bullet$ Two bosons}

We start with the pseudovacuum $\langle\phi \phi|$. Given that the matrix part of the $R$-matrix satisfies $A(\theta_1-\theta_2)=1$ for the boson-boson entry, the individual function $F_{\phi\phi}(m_0,m_1,m_2;\theta_0,\theta_1,\theta_2)$ can be set equal to $1$. This is reproduced by the off-shell Bethe ansatz, because the pseudovacuum simply comes with a coefficient $1$ in the construction. In the case of the pseudovacuum, moreover, ${\cal{I}}_0$ is the identity operator, hence we set $G_{\phi\phi}(\theta_1,\theta_2)=1$. In this case the operator could naturally be a boson, and in particular the boost axiom will work for a scalar function and it will certainly be satisfied by our choice.

{\it $\bullet$ Two fermions}

Let us continue with the state $\langle\psi \psi|$. The $R$-matrix entry for two fermions is $G(\theta_1-\theta_2)$, and the off-shell Bethe ansatz provides us with the appropriate function. Let us set 
\begin{eqnarray}
f(\theta_0,\theta_1,\theta_2)_{m_0,m_1,m_2} = - (B_{10}E_{20}E_{10}D_{20}+E_{10}A_{20}F_{10}E_{20})
\end{eqnarray}
from (\ref{ff}) - mindful that the state $\langle\psi \psi|$ in the algebraic Bethe ansatz contributes to the form factor of two fermions. This evaluates explicitly to
\begin{eqnarray}
f(\theta_0,\theta_1,\theta_2)_{m_0,m_1,m_2} = \frac{2i e^{\frac{i\pi m_0}{k} + \frac{4\theta_0 +\theta_1+\theta_2}{2}}\Big(e^{\frac{2im_0 \pi}{k}}-1\Big)^2 \Big(e^{\theta_1}-e^{\frac{i(m_1+m_2)\pi}{k}+\theta_2}\Big)\sqrt{\sin \frac{m_1 \pi}{k} \, \sin \frac{m_2 \pi}{k}}}{\Big( e^{\frac{i (m_0+m_1) \pi}{k}+\theta_1}-e^{\theta_0}\Big)^2\Big( e^{\frac{i (m_0+m_2) \pi}{k}+\theta_2}-e^{\theta_0}\Big)^2}.
\end{eqnarray}
It can be checked that this function satisfies
\begin{eqnarray}
f(\theta_0,\theta_1,\theta_2)_{m_0,m_1,m_2} = G(m_1,m_2;\theta_1-\theta_2) f(\theta_0,\theta_2,\theta_1)_{m_0,m_2,m_1}.\label{perm3}
\end{eqnarray}
Therefore, if we bear in mind that the $S$-matrix entry for two fermions is $+G(\theta_1 - \theta_2)$ whereas the $R$-matrix entry is $-G(\theta_1 - \theta_2)$, we could set
\begin{eqnarray}
F_{\psi\psi}(m_0,m_1,m_2;\theta_0,\theta_1,\theta_2) = f(\theta_0,\theta_1,\theta_2)_{m_0,m_1,m_2},
\end{eqnarray}
which by virtue of (\ref{perm3}) would satisfy the permutation axiom for all values of $\theta_0 \in \mathbbmss{C}$ (in this sense one says that the Bethe ansatz is {\it off-shell}). In reality, we will not use this expression for the two-fermion form factor, since it turns out not to have good periodicity properties. 

Instead, to respect periodicity, it is be better to use the Karowski-Weisz theorem again. We notice that the fermion-fermion entry of the $S$-matrix is
\begin{eqnarray}
G(\theta) = [m_1+m_2]^{-1}_{\theta},
\end{eqnarray}
therefore we satisfy the axioms by setting the fermion-fermion contribution to
\begin{eqnarray}
G_{\psi\psi}(\theta_1,\theta_2) = \, f^{-1}_{pre}(\theta_1-\theta_2,m_1+m_2),
\end{eqnarray}
where the operator could naturally be taken to be bosonic.

{\it $\bullet$ One boson, one fermion}

We can now deal with the form factors that do mix. As commented above (\ref{evenif}), we can restrict to $m \in \{1,2,...,k-1\}$ so that all the signs are positive. We also call $m_0$ the mass of the auxiliary particle $\theta_0$ in the algebraic Bethe ansatz. From the action of one single $\hat{C}$ operator given in (\ref{numi}) we define
\begin{eqnarray}
f_1(\theta_0,\theta_1,\theta_2)_{m_0,m_1,m_2} =  -E_{10}A_{20} = 2i\frac{e^{\frac{i m_0 \pi}{k}+\frac{\theta_0 + \theta_1}{2}}\sqrt{\sin \frac{\pi m_0}{k} \sin \frac{\pi m_1}{k} }}{e^{\theta_0}-e^{\frac{i\pi (m_0+m_1)}{k}+\theta_1}},
\end{eqnarray}
and
\begin{eqnarray}
f_2(\theta_0,\theta_1,\theta_2)_{m_0,m_1,m_2} = -B_{10}E_{20} = -2i \frac{ e^{ \frac{i m_0 \pi}{k}+\frac{\theta_0 + \theta_2}{2}}\Big(e^{\frac{i m_0 \pi}{k}+\theta_1} - e^{\frac{i m_1 \pi}{k}+\theta_0}\Big)\sqrt{\sin \frac{\pi m_0}{k} \sin \frac{\pi m_2}{k} }}{\Big(e^{\frac{i\pi (m_0+m_1)}{k}+\theta_1}-e^{\theta_0}\Big)\Big(e^{\frac{i\pi (m_0+m_2)}{k}+\theta_2}-e^{\theta_0}\Big)}.
\end{eqnarray}
It is then possible to prove by brute force that
\begin{eqnarray}
f_2(\theta_0,\theta_1,\theta_2)_{m_0,m_1,m_2} = C(m_1,m_2;\theta_{12}) f_2(\theta_0,\theta_2,\theta_1)_{m_0,m_2,m_1} + B(m_1,m_2;\theta_{12})f_1(\theta_0,\theta_2,\theta_1)_{m_0,m_2,m_1}\nonumber\\\label{perm1}
\end{eqnarray}
and
\begin{eqnarray}
f_1(\theta_0,\theta_1,\theta_2)_{m_0,m_1,m_2} = E(m_1,m_2;\theta_{12}) f_1(\theta_0,\theta_2,\theta_1)_{m_0,m_2,m_1} + D(m_1,m_2;\theta_{12})f_2(\theta_0,\theta_2,\theta_1)_{m_0,m_2,m_1},\nonumber\\\label{perm2}
\end{eqnarray}
where we will often use the notation 
\begin{eqnarray}
\theta_{ij} = \theta_i - \theta_j
\end{eqnarray}
throughout, where $i\neq j$ could be any of $0$, $1$ or $2$.

These two conditions reconstruct the permutation axiom applied to the two individual functions that mix: $F_{\phi\psi}$ and $F_{\psi\phi}$. Specifically we can set
\begin{eqnarray}
F_{\phi\psi}(m_0,m_1,m_2;\theta_0,\theta_1,\theta_2) =  f_2(\theta_0,\theta_1,\theta_2)_{m_0,m_1,m_2}\label{a}
\end{eqnarray}
and
\begin{eqnarray}
F_{\psi\phi}(m_0,m_1,m_2;\theta_0,\theta_1,\theta_2) =  f_1(\theta_0,\theta_1,\theta_2)_{m_0,m_1,m_2}\label{b},
\end{eqnarray}
again for any $\theta_0 \in \mathbbmss{C}$ ({\it off -shell}). This shows that (\ref{perm1}) and (\ref{perm2}) perfectly reproduce the permutation axiom, since $C$ is the $S$-matrix entry for $\phi\psi \to \phi\psi$, $D$ is the $S$-matrix entry for $\psi\phi \to \phi\psi$, $+E$ is the $S$-matrix entry for $\psi\phi \to \psi\phi$ and $B$ is the $S$-matrix entry for $\phi\psi \to \psi\phi$. 

In order to reproduce the method devised in \cite{BabuF}, we now need to assemble a few components. First, let us define
\begin{eqnarray}
&&\tilde{\Phi}_{m_0,m_i}(\theta_{i0}) \equiv \frac{1}{F^{-1}(m_0,m_i;\theta_{i0}) \, F^{-1}(k-m_0,m_i;\theta_{i0}+i\pi)}, \qquad \theta_{i0}=\theta_i - \theta_0.
\end{eqnarray}
The advantage of this definition resides in its transformation properties under periodicity: let us start with $i=1$ and shift $\theta_1$ by $+2i \pi$. To begin with, we have checked that braiding unitarity, crossing symmetry and analiticity also hold for $\sigma^{-1}$, and not only for $\sigma^{-2}$, if one halves the powers everywhere accordingly. Likewise we have checked that the the relations satisfied by $F^{-1}(m_1,m_2;\theta)$ can be obtained from those satisfied by $F^{-2}(m_1,m_2;\theta)$ by halving the powers everywhere - analyticity also being preserved (no square roots appearing anywhere). We therefore have
\begin{eqnarray}
&&\tilde{\Phi}_{m_0,m_1}(\theta_1 + 2 i \pi - \theta_0) \nonumber\\
&& =\frac{1}{F^{-1}(m_0,m_1;\theta_1+2 i \pi-\theta_0)F^{-1}(k-m_0,m_1;\theta_1+3 i \pi - \theta_0)} \nonumber\\
&&= \frac{1}{F^{-1}(m_0,m_1;-\theta_1+\theta_0)F^{-1}(k-m_0,m_1;-\theta_1+\theta_0- i \pi)} \nonumber\\
&&=\frac{\sigma^{-1}(m_0,m_1;\theta_1-\theta_0) \sigma^{-1}(k-m_0,m_1;\theta_1-\theta_0+i\pi)}{F^{-1}(m_0,m_1;\theta_1-\theta_0)F^{-1}(k-m_0,m_1;\theta_1- \theta_0+i \pi)}\nonumber\\
&&=\tilde{\Phi}_{m_0,m_1}(\theta_1-\theta_0) \times f_{m_0,m_1}(\theta_1-\theta_0),\label{tri1}
\end{eqnarray} 
having in addition used the crossing properties of the dressing factor. We have also used the symmetry under exchange of $m_1$ and $m_2$ of $\sigma^{-1}(m_1,m_2;\theta)$ and $F^{-1}(m_1,m_2;\theta)$. The function $f_{m_1,m_2}(\theta)$ is given by \cite{Frolov:2023lwd}
\begin{eqnarray}
f_{m_1,m_2}(\theta) = \frac{\sinh \Big(\frac{\theta}{2} - \frac{i\pi}{2k}(m_1-m_2)\Big)}{\sinh \Big(\frac{\theta}{2} - \frac{i\pi}{2k}(m_1+m_2)\Big)}.\label{prop}
\end{eqnarray}
Likewise, if we focus on $i=2$ and shift $\theta_2$ by $-2i\pi$ (the reason for which will shortly become apparent), we have
\begin{eqnarray}
&&\tilde{\Phi}_{m_0,m_2}(\theta_2 - 2 i \pi - \theta_0) \nonumber\\
&& =\frac{1}{F^{-1}(m_0,m_2;\theta_2-2 i \pi-\theta_0)F^{-1}(k-m_0,m_2;\theta_2- i \pi - \theta_0)}\nonumber\\
&&=\frac{1}{\sigma^{-1}(m_0,m_2;\theta_{20}-2i\pi)F^{-1}(m_0,m_2;2i\pi - \theta_{20})\sigma^{-1}(k-m_0,m_2;\theta_{20}-i\pi)F^{-1}(k-m_0,m_2;i\pi - \theta_{20})}\nonumber\\
&&\frac{1}{f_{m_0,m_2}(\theta_{20}-2i\pi)F^{-1}(m_0,m_2; \theta_{20})F^{-1}(k-m_0,m_2;i\pi + \theta_{20})}\nonumber\\
&&=f^{-1}_{m_0,m_2}(\theta_{20}) \tilde{\Phi}_{m_0,m_2}(\theta_2-\theta_0),\label{tri2}
\end{eqnarray}
having used at the very end the properties of the function $f_{m_1,m_2}(\theta)$ in (\ref{prop}).

By brute force computation one can also check that
\begin{eqnarray}
F_{\psi\phi}(m_0,m_1,m_2;\theta_0,\theta_1,\theta_2-2i\pi)=f^{-1}_{m_0,m_2}(\theta_{02}) F_{\phi\psi}(m_0,m_2,m_1;\theta_0,\theta_{2},\theta_1)
\end{eqnarray}
and
\begin{eqnarray}
F_{\phi\psi}(m_0,m_1,m_2;\theta_0,\theta_1+2i\pi,\theta_2)=f_{m_0,m_1}(\theta_{01}) F_{\psi\phi}(m_0,m_2,m_1;\theta_0,\theta_{2},\theta_1).
\end{eqnarray}

Finally, we have
\begin{eqnarray}
&&\sigma^{-1}(m_0,m_1;\theta_{10}+2i\pi) =\frac{f_{k-m_0,m_1}(\theta_{10}+i\pi)}{\sigma^{-1}(k-m_0,m_1;\theta_{10}+i\pi)} \nonumber\\
&&=\frac{f_{k-m_0,m_1}(\theta_{10}+i\pi)}{f_{m_0,m_1}(\theta_{10})}\sigma^{-1}(m_0,m_1;\theta_{10}) = \frac{\sigma^{-1}(m_0,m_1;\theta_{10})}{f_{m_0,m_1}(\theta_{10})f_{m_0,m_1}(\theta_{01})},
\end{eqnarray}
and also (using the symmetry of $\sigma(m_1,m_2;\theta)$ under $m_1 \leftrightarrow m_2$ and its braiding-unitarity property)
\begin{eqnarray}
&&\sigma^{-1}(m_0,m_2;\theta_{20}-2i\pi) =\frac{1}{\sigma^{-1}(m_0,m_2;2i\pi+\theta_{02})}=\frac{\sigma^{-1}(k-m_0,m_2;i\pi+\theta_{02})}{f_{k-m_0,m_2}(i\pi+\theta_{02})}=\nonumber\\
&&\frac{f_{m_0,m_2}(\theta_{02})}{f_{k-m_0,m_2}(i\pi+\theta_{02})}\frac{1}{\sigma^{-1}(m_0,m_2;\theta_{02})}=f_{m_0,m_2}(\theta_{02})f_{m_0,m_2}(\theta_{20})\sigma^{-1}(m_0,m_2;\theta_{20}).
\end{eqnarray}

We can now assemble all the components and obtain the form factors satisfying the axioms. We define the fermion-boson individual function as
\begin{eqnarray}
&&G_{\psi\phi}(\theta_1,\theta_2)=\nonumber\\
&&\int_{C_{12}} \frac{d\theta_0  \, e^{H_{12}}}{\prod_{i=1}^2 F^{-1}(m_0,m_i;\theta_{i0})F^{-1}(k-m_0,m_i;\theta_{i0+i\pi})} F_{\psi\phi}(m_0,m_1,m_2;\theta_0,\theta_1,\theta_2) \prod_{i=1}^2 \sigma^{-1}(m_0,m_i;\theta_{i0}),\nonumber\\ \label{inte1}
\end{eqnarray}
and the boson-fermion one as 
\begin{eqnarray}
&&G_{\phi\psi}(\theta_1,\theta_2)=\nonumber\\
&&\int_{C_{12}} \frac{d\theta_0 \, e^{H_{12}}}{\prod_{i=1}^2 F^{-1}(m_0,m_i;\theta_{i0})F^{-1}(k-m_0,m_i;\theta_{i0+i\pi})} F_{\phi\psi}(m_0,m_1,m_2;\theta_0,\theta_1,\theta_2) \prod_{i=1}^2 \sigma^{-1}(m_0,m_i;\theta_{i0}).\nonumber\\\label{inte2}
\end{eqnarray}
If the operator, of which we are computing the form factor, has spin $s$, we define
\begin{eqnarray}
H_{12} = (\theta_1+\theta_2) + (s-2)\theta_0,\label{h}
\end{eqnarray}
to satisfy the boost axiom. This turns out to produce the naively expected sign under periodicity for an operator with half-integer spin $s$.

To prove periodicity, we shift $\theta_1 + 2 i \pi$ in the boson-fermion formula. In the fermion-boson formula instead, we shift $\theta_1+2i\pi$ but simultaneously shift the integration variable $\theta_0 + 2 \pi i$, so as a net effect this is equivalent to shift $\theta_2 - 2 i \pi$ in the integrand. $H_{12}$ is a symmetric function under the exchange of $1$ and $2$, so that permutation is not spoilt. The integration contour will also be chosen to be symmetric under the exchange of $1$ and $2$, so that permutation is not spoilt.

\begin{figure}
\centerline{\includegraphics[width=18cm]{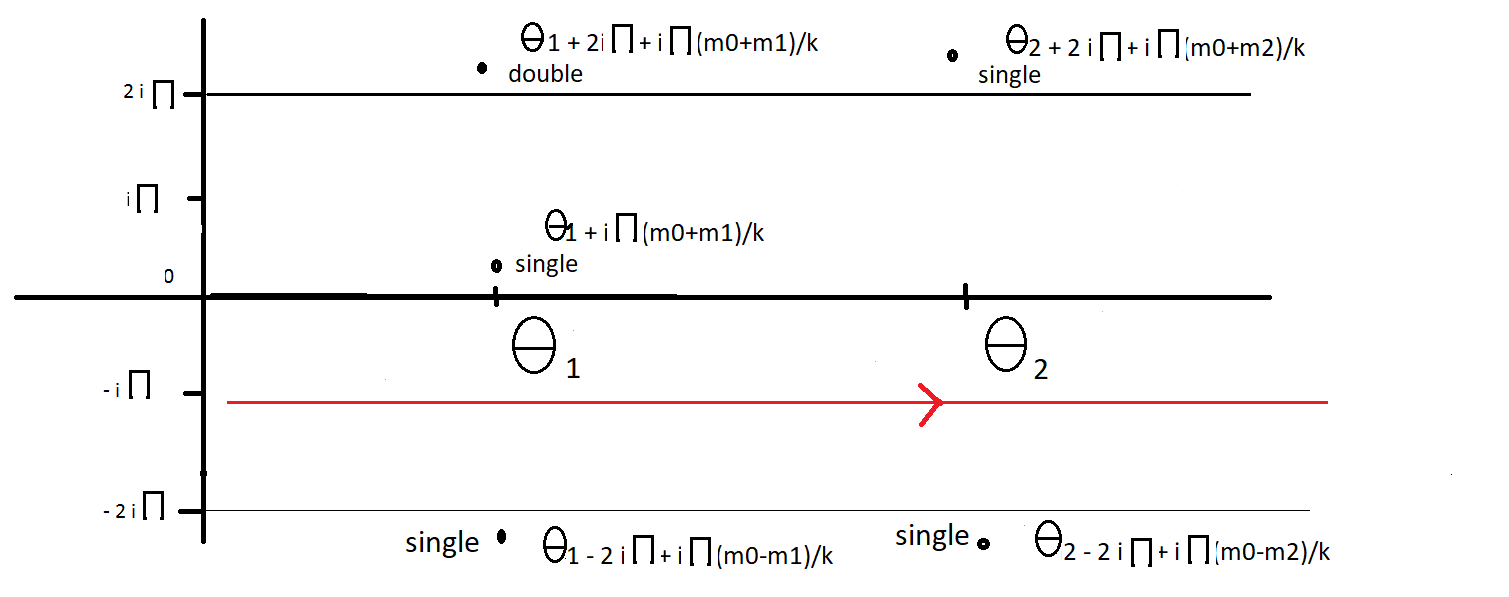}}
\caption{The locations of the poles of the fermion-boson integrand are denoted with little circular dots with their respective order, in a portion of the complex plane of $\theta_0$. We have assumed $0<m_0<min[m_1,m_2]$ and $k$ sufficiently large. In red we have indicated the integration countour.}
\label{fig1}
\end{figure}

\begin{figure}
\centerline{\includegraphics[width=18cm]{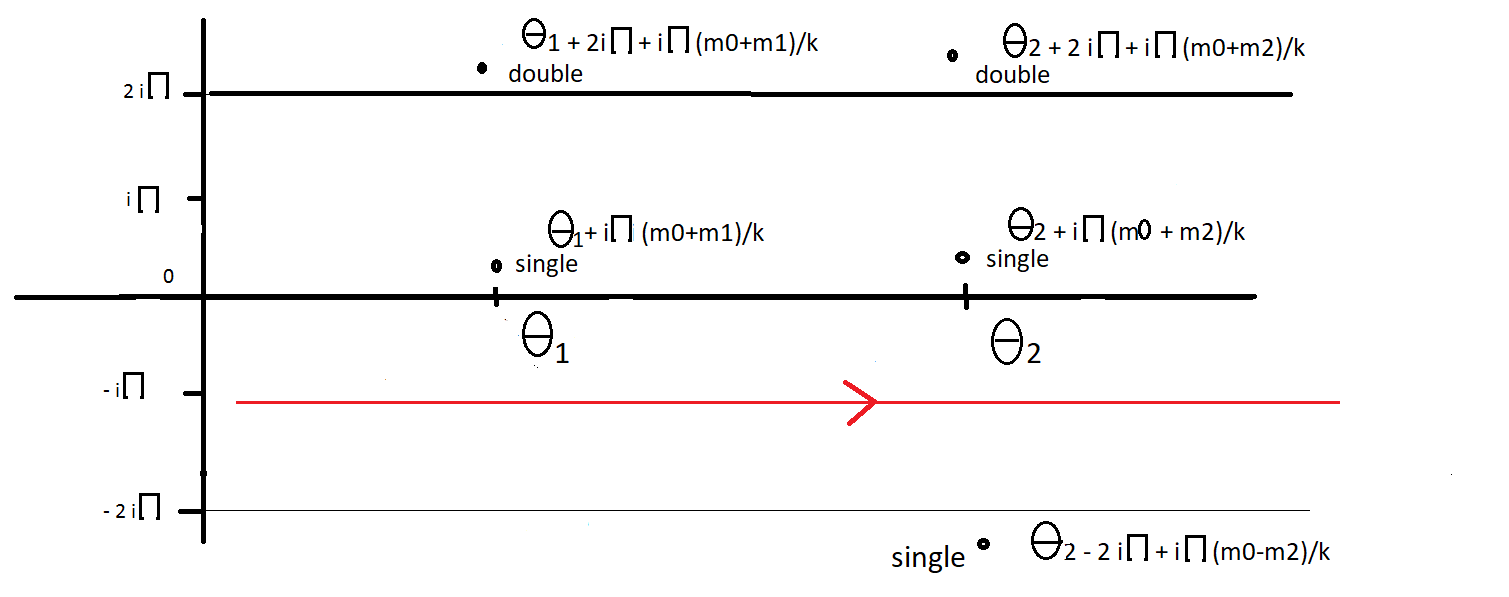}}
\caption{The locations of the poles of the boson-fermion integrand are denoted with little circular dots with their respective order, in a portion of the complex plane of $\theta_0$. We have assumed $0<m_0<min[m_1,m_2]$ and $k$ sufficiently large. In red we have indicated the integration countour.}
\label{fig2}
\end{figure}

If we focus on the strip Im$(\theta_0) \in [-2\pi,2\pi]$, a careful analysis of poles and zeros of the integrand of (\ref{inte1}) and (\ref{inte2}) reveals the picture in Fig. \ref{fig1} and Fig. \ref{fig2}.

We will chose the contour to run through the points $\theta_1- i \pi - i \epsilon$ and $\theta_2 - i \pi - i \epsilon$ as shown in the figures.  

Periodicity works this way:
\begin{itemize}
\item for the boson-fermion form-factor, we shift $\theta_1 + 2 i \pi$. By virtue of the formulas obtained above, the integrand becomes identical to the integrand of the fermion-boson form-factor.  
\item for the fermion-boson form-factor, we shift $\theta_1 + 2 i \pi$ and change integration $\theta_0 + 2 i \pi$. By virtue of the formulas obtained above, the integrand becomes identical to the integrand of the fermion-boson form-factor. The contour shifts but it encounters no poles as it settles back into its original shape.
\end{itemize}

To complete all cases, we can go back to the two-fermion form factor. 

{\it $\bullet$ Recombining the two copies}

One ought to tensor together two copies of the integral according to the pattern of states, and the total should then be supplemented by a factor
\begin{eqnarray}
F^{-2}(m_1,m_2;\theta_{12})F^{total}_{pre}(m_1,m_2;\theta_{12}),
\end{eqnarray}
so that the complete form factor will be
\begin{eqnarray}
F^{-2}(m_1,m_2;\theta_{12})F^{total}_{pre}(m_1,m_2;\theta_{12}) \, G_{ac}(\theta_1,\theta_2) G_{bd}(\theta_1,\theta_2). 
\end{eqnarray}

We notice that we seem to find the axioms to be satisfied for any value of $m_0$ (not necessarily integer). We have not explored whether this is a genuine degree of freedom in our solution, or whether it will be fixed in some way by the theory.

Let us remark that the kinematical singularity axiom degenerates for the case of two particles, as there are no strict sub-channels\footnote{We can notice in passing that our integrals display no singularity when $\theta_1 = \theta_2 + i \pi$ (no poles of the integrands pinching the contour), therefore we technically fulfil a degenerate version of the kinematical singularity axiom with $0$ on the right hand side.}.

The issue related to the bound state axiom is something which will require a more detailed knowledge of the operator. A naive reading of the bound state form factor axiom implies that the residue at the bound state pole must be proportional to the one point form factor of the operator $\cal{O}$ with the bound state creation operator. It may be necessary to suitably modify our expressions to account for the bound state singularities \cite{DelfinoMussSimo} - in fact we cannot make any claim about the uniqueness of the formulas which we have so far constructed. We reserve the investigation of bound states for future work.  

\section{Convergence}

We can compute the asymptotics of the components of our integrals (for generic value of the parameters): if we rescale 
\begin{eqnarray}
x \equiv \frac{y}{|\mbox{Re} \, \theta|},
\end{eqnarray}
and subsequently send $|\mbox{Re} \, \theta| \to \infty$, we can expand the integrand first and then perform the integrals. We obtain
\begin{eqnarray}
&&\sigma^{-2}(m_1,m_2,\theta) \sim \exp (\pm 1) \int_0^\infty \frac{dy}{y} \, \Big(-\frac{4m_1 m_2}{k^2}\Big) \sinh \frac{y}{i\pi} = \exp \Big(\pm \frac{2i \pi \, m_1 m_2}{k^2}\Big), \qquad \mbox{Re} \, \theta \longrightarrow \pm \infty,\nonumber\\
&&F^{-2}(m_1,m_2,\theta) \sim \exp \int_0^\infty \frac{dy}{y} \, \Big(\frac{2 m_1 m_2 |\mbox{Re} \, \theta|}{k^2 y}\Big)\Big(\cos\frac{y}{\pi}-1\Big) = \exp \Big(\frac{-m_1 m_2 |\mbox{Re} \, \theta|}{k^2}\Big), \quad \mbox{Re} \, \theta \longrightarrow \pm \infty.\nonumber
\end{eqnarray}
However we should point out that this expansion is somewhat dangerous, since the subleading order in $|\mbox{Re} \, \theta|^{-1}$ of the second line turns out to be multiplied by a divergent integral. The conclusions which we draw are therefore only naive and a more thorough analysis might be required. Numerical experimentation reveals that within the error bars a very good approximation is obtained by the following formula:
\begin{eqnarray}
&&F^{-2}(m_1,m_2,\theta) \sim \exp \Big( \frac{\mp m_1 m_2 (\theta - i \pi)}{k^2} \Big), \qquad \mbox{Re} \, \theta \to \pm \infty,
\end{eqnarray}
while numerically the asymptotics of $\sigma^{-2}$ are quite well reproduced by the formula we gave above.

By inspecting the integrals we see that the naive asymptotic behaviour is therefore
\begin{eqnarray}
&&G_{\phi\psi} \sim G_{\psi\phi} \sim \int^{+\infty} d(\mbox{Re} \, \theta_0) \, \, e^{\mbox{Re} \, \theta_0 \times (s - \frac{5}{2}+\frac{m_1+m_2}{2k})}, \qquad \mbox{Re} \, \theta_0 \to +\infty,\nonumber\\
&&G_{\phi\psi} \sim G_{\psi\phi} \sim \int_{-\infty} d(\mbox{Re} \, \theta_0) \, \, e^{\mbox{Re} \, \theta_0 \times (s - \frac{3}{2}+\frac{m_1+m_2}{2k})}, \qquad \mbox{Re} \, \theta_0 \to -\infty,
\end{eqnarray}
hence we get convergence in the region
\begin{eqnarray}
s \in \Big(\frac{3}{2}+\frac{m_1+m_2}{2k}, \frac{5}{2}-\frac{m_1+m_2}{2k}\Big).
\end{eqnarray}
If we wish to set the spin to $s=\frac{1}{2}$ or $s=\frac{3}{2}$ for instance, we might consider analytically continue the integrals either in $m_1,m_2$ or $k$ to a region of convergence (for instance negative values of $\frac{m_1+m_2}{k}$) - see for instance the discussions in \cite{conve}. Another option would be to modify the contour - our choice in figures \ref{fig1} and \ref{fig2} was the simplest choice compatible with the axioms, but there could be more elaborated choices (see for example \cite{BabuF}) which may produce a final result in a wider region of the parameters. We have not explored these possibilities, and we plan to come back to this issue in future work.

Another worthwhile consideration is that our $H_{12}$ function (\ref{h}) is tuned to a standard assignement of statistics - equivalently, to the statistics of the bare states. As shown in \cite{Frolov:2023lwd,gamma2}, the coproduct acquires a deformation. This alters the basic statistics \cite{Frolov:2023lwd}, and therefore one might need to revisit our assignment of $H_{12}$ depending on the operators considered. This in turn might change the convergence properties of the integral, which is extremely sensitive to the exponential $e^{H_{12}}$.    

The occurrence of divergences
in the integrals, both in the off-shell Bethe ansatz method and in the Lukyanov method, are in fact contemplated and seem to be difficult to avoid. It appears that the systematic
analytic continuations of the integral representations has to be implemented. In the case of the pure RR massless theory an additional source of divergence is due to the infrared limit entangled with the poor convergence of the series expansion, as discussed in \cite{AleForm} - section 8.1 and particularly section 8.1.2. in that paper. 
 

\section{Perturbation theory}

There is limit where formulas simplify - in particular the dressing factor becomes the identity - which is the $k\to \infty$ limit. It turns out that we can expand our formulas in orders $\frac{1}{k}$ by relying on the integral representations, expanding the integrand first, and then performing the integral. We assemble here the various pieces:
\begin{eqnarray}
&&\log \sigma^{-1}(m_1,m_2,\theta) = 1 + \frac{i m_1 m_2 \, \pi (\sinh \theta \, - \, \theta)}{k^2 (\cosh \theta \, - \, 1)}+...,\nonumber\\
&&\log F^{-1}(m_1,m_2,\theta) = 1 + \frac{1}{8k^2} m_1 m_2 \frac{\theta(\theta-2 i \pi)+2i\sinh \theta \, (\pi + i \theta)}{\sinh^2 \theta} +\frac{\pi^2 m_1 m_2}{8 k^2}+...,\nonumber\\
&&\log F^{-1}(k-m_1,m_2,\theta) = 1 +\frac{m_2\big(2+i(\pi + i \theta)\tanh\frac{\theta}{2}\big)}{2k} + \nonumber\\
&&\qquad \qquad \qquad \qquad+ \frac{m_1 m_2\Big(-4 +\frac{(\theta - i \pi)(- i \pi + \theta + 2 \sinh \theta)}{\cosh^2 \frac{\theta}{2}}\Big)}{8k^2}....
\end{eqnarray}
We can also expand $F_{\psi \phi}$ and $F_{\phi\psi}$ up to order $\frac{1}{k^2}$ (their expansion not being very illuminating), and finally the CDD-like pre-factor (in the case of $m_1>m_2$) 
\begin{eqnarray}
&&\Phi(m_1,m_2,\theta) = 1+ \frac{2i\big(m_1(m_2-1)+m_2\big)\pi \coth \frac{\theta}{2}}{k}+...,\nonumber\\ 
&&F^{total}_{pre}(m_1,m_2,\theta) = 1+ \frac{2i\big(m_1(m_2-1)+m_2\big)(\pi +i \theta)\coth \frac{\theta}{2}}{k}+...
\end{eqnarray}

Let us focus for instance on the fermion-boson and boson-fermion form factors. Assembling all the pieces we find that at the first non-trivial order we can naively perform the final contour integration of figures \ref{fig1} and \ref{fig2} if we analytically continue in the spin $s$ to the region $\frac{3}{2}<s<\frac{5}{2}$, with the spin acting as a regulator. We then obtain
\begin{eqnarray}
&&G_{\phi\psi}(\theta_1,\theta_2) = \label{33}\\
&&\quad-\frac{4 \pi \sqrt{m_0 m_2}e^{-i\pi s + \theta_1 + \frac{\theta_2}{2}}}{k} \Bigg[e^{\theta_2} \frac{_2F_1(1,\frac{5}{2}-s,\frac{7}{2}-s,-e^{\theta_2})}{5-2s}+\frac{ _2F_1(1,-\frac{3}{2}-s,-\frac{1}{2}-s,-e^{-\theta_2})}{-3+2s}\Bigg]+ ...,\nonumber
\end{eqnarray}
\begin{eqnarray}
&&G_{\psi\phi}(\theta_1,\theta_2) =  \label{34}\\
&&\quad-\frac{4 \pi \sqrt{m_0 m_1}e^{-i\pi s + \theta_2 + \frac{\theta_1}{2}}}{k} \Bigg[e^{\theta_1} \frac{_2F_1(1,\frac{5}{2}-s,\frac{7}{2}-s,-e^{\theta_1})}{5-2s}+\frac{ _2F_1(1,-\frac{3}{2}-s,-\frac{1}{2}-s,-e^{-\theta_1})}{-3+2s}\Bigg]+ ...\nonumber,
\end{eqnarray} 
We also have
\begin{eqnarray}
G_{\psi\psi}(\theta_1,\theta_2) = 1 - \frac{i(m_1+m_2)(\pi + i \theta_{12})\coth \frac{\theta}{2}}{k}+...,\qquad G_{\phi,\phi}=1+....
\end{eqnarray}
The expressions (\ref{33}) and (\ref{34}) have poles at the values of the spin $s=\frac{1}{2}, \frac{3}{2},...$ (recalling that the operator is necessarily fermionic): 
\begin{eqnarray}
&&G_{\psi\phi} = - \frac{2i\pi \sqrt{m_0 m_2}\, e^{\theta_1 - \frac{\theta_2}{2}}}{k (s-\frac{1}{2})}+..., \qquad G_{\phi\psi} = - \frac{2i\pi \sqrt{m_0 m_1}\, e^{\theta_2 - \frac{\theta_1}{2}}}{k (s-\frac{1}{2})}+..., \qquad s \sim \frac{1}{2},\nonumber\\
&&G_{\psi\phi} = - \frac{2i\pi \sqrt{m_0 m_2}\, e^{\theta_1 + \frac{\theta_2}{2}}}{k (s-\frac{3}{2})}+..., \qquad G_{\phi\psi} = - \frac{2i\pi \sqrt{m_0 m_1}\, e^{\theta_2 + \frac{\theta_1}{2}}}{k (s-\frac{3}{2})}+..., \qquad s \sim \frac{3}{2}.\nonumber
\end{eqnarray}
 
It is amusing to observe that we could tune the - so far unspecified - parameter $m_0$ to neutralise the divergence. We could for instance choose to send $m_0 \sim (s - \frac{1}{2})^2$ as long with $s \to \frac{1}{2}$, and we would probably be safe in the limit (the same with $\frac{3}{2}$). 

We can observe that, if we have anywhere a $G_{\phi\psi}$ or a $G_{\psi\phi}$, such as in the form factor of the states $(\phi\hat{\otimes}\phi)\otimes (\psi\hat{\otimes}\phi)$ for instance, then the leading order of the form factor is dictated by (\ref{33}) or (\ref{34}), for instance
\begin{eqnarray}
\langle 0 |{\cal{O}}|(\phi\hat{\otimes}\phi)_{\theta_1}\otimes (\psi\hat{\otimes}\phi)_{\theta_2}\rangle =  \mbox{``}\Phi \, F^{-2} \, G_{\phi\psi}G_{\phi\phi}\mbox{"}= (\ref{33})+ ...
\end{eqnarray}
(with ${\cal{O}}$ being fermionic). Similarly for other combinations involving $G_{\phi\psi}$ or $G_{\psi\phi}$.

\section{Conclusions}

In this paper we have studied the form factors of the mixed flux $AdS_3 \times S^3 \times T^4$ scattering theory in the relativistic limit, using the formulation of \cite{Frolov:2023lwd}, which uses a distinct limit from the one studied in \cite{gamma2}. Here we have focused on the massive sector and derived the minimal solutions to the form factor axioms corresponding to two-particles. We have then constructed the complete set of two particle form factors using the off-shell Bethe ansatz method of \cite{BabuF}.

The highlight of our construction is that we can solve the axioms of two-particle form factors, although the integrals for the fermion-boson and boson-fermion case appears to be divergent for certain value of the spin $s$. We have discussed possibilities for regularising the integral, for instance analytically continuing in the parameter $\frac{1}{k}$ which appears to behave like a small coupling in perturbation theory. It is also possible that the contour used for the integrals could be modified. Finally, the deformed statistics of the dressed particles might force one to change the exponential function $H_{12}$, upon which the convergence of the integrals depends quite strongly. This will have to be done in accordance with which operator one is considering.

There are a number of directions which one should explore, most importantly whether one can properly cure the divergence. As always, it would be extremely useful to obtain perturbative checks with Feynman diagrams to show that we are on the right track. Generalising to more particles appears in principle standardised by the off-shell Bethe ansatz method, but not too easily implementable practically, as it appears to depend quite significantly on the precise distribution of singularities of the integrand. In particular the bound state axiom and the kinematical singularities would have to be carefully analysed. It would also be interesting to approach the problem using Lukyanov's method \cite{Lukyanov,fofa} and see whether we obtain a different result. We plan to come back to these issues in future work.

\section*{Acknowledgements}
The author greatly benefited in exchanges with Frolov, Polvara and Sfondrini and Fontanella, Ohlsson Sax and Stefa\'nski in establishing the convergence of views on special cases of the dressing factor. The author thanks Davide Polvara for extremely interesting discussions and offline useful explanations of the paper \cite{Frolov:2023lwd} at the {\it Integrability, Dualities and Deformations} meeting in Durham, 2023. The authour thanks Ohlsson Sax, Riabchenko and Stefa\'nski for very kindly sharing their manuscript \cite{BOD} prior to publication, and also for their reciprocally reading of this paper. The author has vastly benefited from conversation, annotations on this paper and numerous points raised by Bogdan Stefa\'nski - the email exchange is truly appreciated. The author thanks Juan Miguel Nieto Garc\'ia for reading the manuscript and providing very useful suggestions. The author thanks the Isaac Newton Institute in Cambridge and the organisers of the programme {\it Machine Learning Toolkits and Integrability Techniques in Gravity}, EPSRC grant nr. EP/R014604/1, for their kind support and hospitality. This work has been
supported by EPSRC-SFI under the grant EP/S020888/1 {\it Solving Spins and Strings}.

\section*{\label{sec:Data}Data Access Statement}

No data beyond those presented in this paper are needed to validate its results.

\end{document}